\documentclass[9pt,twocolumn,twoside]{pnas}
\templatetype{pnas}

\usepackage[utf8]{inputenc} 
\usepackage[T1]{fontenc}  
\usepackage{hyperref}    
\usepackage{url}      
\usepackage{booktabs}    
\usepackage{amsfonts}    
\usepackage{nicefrac}    
\usepackage{microtype}   
\usepackage{lipsum}
\usepackage{graphicx}
\usepackage[utf8]{inputenc}
\usepackage[english]{babel}
\usepackage{amsmath}

\usepackage{natbib}
\usepackage{sidecap}
\usepackage{subfig}

\definecolor{brickred}{rgb}{0.79, 0.25, 0.32}
\newcommand{\charlotte}[1]{{\color{black}#1}}
\newcommand{\alex}[1]{{\color{black}#1}}

\newcommand{\jr}[1]{{\color{black}#1}}

\title{Deep language models are poor long-range predictors compared to the brain}
\title{The human brain better anticipate the distant representations of speech than current language models}
\title{Short-sighted AI: Brain imaging reveals the inability of deep language models to predict the long-range future}

\title{Deep Language Algorithms Lack the Long-range Forecasts of the Human Brain}
\title{Deep Language Algorithms are poor long-range predictors compared to the human Brain}
\title{Deep Language Algorithms reveal the long-range and abstract predictions of the brain}
\title{Deep learning reveal the deep and distant language predictions of the brain}

\title{Deep Language Algorithms reveal the long-range Forecasts of the Human Brain}
\title{Deep Language Algorithms reveal the superiority of the brain for long-range forecasts}

\title{Deep Language Algorithms Lack the Long-range Forecasts of the Human Brain}

\title{Language predictions: fundamental differences between brains and machines}
\title{Language predictions in brains and machines}
\title{Different speech predictions in brains and machines}
\title{Long-range and hierarchical speech prediction in brains and algorithms}
\title{Different speech prediction in brains and algorithms}
\title{Deep and distant language predictions in brains and algorithms}

\title{Long-range and hierarchical language predictions in brains and algorithms}

\author[1,2]{Charlotte Caucheteux}
\author[2]{Alexandre Gramfort}
\author[1,3]{Jean-Rémi King}

\affil[1]{Facebook AI Research, Paris, France}
\affil[2]{Université Paris-Saclay, Inria, CEA, Palaiseau, France}
\affil[3]{École normale supérieure, PSL University, CNRS, Paris, France}


\begin{abstract}

Deep learning has recently made remarkable progress in natural language processing. Yet, the resulting algorithms remain far from competing with the language abilities of the human brain. Predictive coding theory offers a potential explanation to this discrepancy: while deep language algorithms are optimized to predict adjacent words, the human brain would be tuned to make long-range and hierarchical predictions. To test this hypothesis, we analyze the fMRI brain signals of 304 subjects each listening to $\approx$70\,min of short stories. After confirming that the activations of deep language algorithms linearly map onto those of the brain, we show that enhancing these models with long-range forecast representations improves their brain-mapping. The results further reveal a hierarchy of predictions in the brain, whereby the fronto-parietal cortices forecast more abstract and more distant representations than the temporal cortices. Overall, this study strengthens predictive coding theory and suggests a critical role of long-range and hierarchical predictions in natural language processing.

\end{abstract}

\begin{document} 

\maketitle
\ifthenelse{\boolean{shortarticle}}{\ifthenelse{\boolean{singlecolumn}}{\abscontentformatted}{\abscontent}}{}



\dropcap{I}n less than three years, deep learning has made considerable progress in text generation, translation and completion~\citep{vaswani_attention_2017,radford_language_nodate, brown_language_2020,fan_hierarchical_2018} thanks to algorithms trained with a simple learning rule: predicting words from their adjacent context.
%
Remarkably, the activations of these models have been shown to linearly map onto human brain responses to speech and text~\citep{jain_incorporating_2018,toneva_interpreting_2019, caucheteux_language_2020,schrimpf_artificial_2020, goldstein_thinking_2021}
\jr{. Besides, }this mapping appears to primarily depend on the algorithms' ability to predict future words~\citep{caucheteux_language_2020,schrimpf_artificial_2020}\jr{, hence suggesting that this learning rule suffices to make them converge to brain-like computations}.

Yet, \jr{a major gap remains} between humans and these algorithms: current language models are still poor at story generation and summarization as well as dialogue and question answering~\citep{holtzman_curious_2020, wiseman_challenges_2017, thakur_beir_2021, raffel_exploring_2020, krishna_hurdles_2021}; they fail to capture many syntactic constructs and semantics properties~\citep{lakretz_emergence_2019, arehalli2020neural, lakretz_can_2021, baroni_linguistic_2020, lake_word_2021}, and their \jr{linguistic} 
understanding is often superficial~\citep{marcus2020gpt2, lake_word_2021,baroni_linguistic_2020, arehalli2020neural}. 



\begin{figure*}[ht]
\centering
\includegraphics[width=\linewidth]{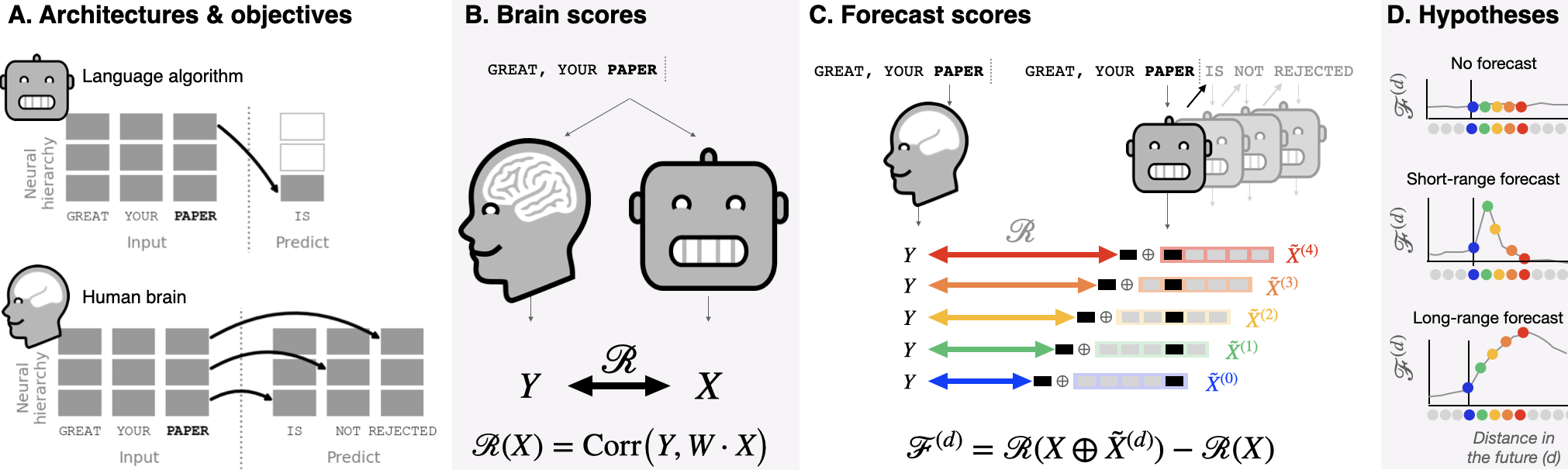}
\caption{
\jr{
\textbf{Approach.}
\textbf{A.}
Deep language algorithms are typically trained to predict words from their close contexts.
Unlike these algorithms, the brain makes, according to predictive coding theory, \emph{(i)} long-range and \emph{(ii)} hierarchical predictions.
\textbf{B.} 
To test this hypothesis, we first extract the fMRI signals of 304 subjects each listening to $\approx$70\,min of short stories ($Y$) as well as the activations of a deep language algorithm ($X$) input with the same stories. 
We then quantify the similarity between $X$ and $Y$ with a ``brain score”: a Pearson correlation $\mathcal{R}$ after an optimal linear projection $W$ (Methods \ref{methods-brainscore}).
\textbf{C.} 
To test whether adding representations of future (or predicted, see Figure \ref{fig:controls}) words improves this correlation, we concatenate ($\oplus$) the network's activations ($X$, depicted here as a black rectangle) to the activations of a “forecast window” ($\tilde{X}$, depicted here as a colored rectangle). 
We use principal component analysis to reduce the dimensionality of the forecast window down to the dimensionality of $X$. 
Finally, $\mathcal{F}$ quantifies the gain of brain score obtained by enhancing the activations of the language algorithm to this forecast window.
We repeat this analysis with variably distant windows ($d$, Methods \ref{methods-forecast-window}).
\textbf{D.} 
A flat forecast score across distances would indicate that forecast representations do not make the algorithm more similar to the brain (top). 
By contrast, a forecast score peaking at $d>1$ (bottom) would indicate that the model lacks brain-like forecast.
The peak of $\mathcal{F}^d$ indicates how far off in the future the algorithm would need to forecast representations to be most similar to the brain.
}
}
\label{fig:fig1}
\end{figure*}

Predictive coding theory~\citep{rumelhart_interactive_1982, rao_predictive_1999, friston_predictive_2009} offers a potential explanation to these shortcomings: while deep language models are tuned to predict the very next word, this theory suggests that the human brain predicts \emph{(i)} long-range and \emph{(ii)} hierarchical representations (Figure \ref{fig:fig1}A). Previous work has already evidenced 
\jr{speech} predictions in the brain, by correlating 
word \jr{or phonetic} surprisal
with functional Magnetic Resonance Imaging (fMRI)
\citep{willems_prediction_2016, lopopolo_using_2017, okada_neural_2018, shain_fmri_2019}, electroencephalography~\citep{heilbron_tracking_2019, donhauser_two_2020}, magnetoencephalography~\citep{mousavi_brain_2020} and  electrocorticography
~\citep{forseth_language_2020, goldstein_thinking_2021}. However, such surprisal estimates 
\jr{derive from models trained to predict} 
the very next \jr{token (i.e. word \emph{or} phoneme)}, \charlotte{and reduce down their output to a single number: the probability of the next token.}
%
\charlotte{Consequently, \emph{(i)} the nature of the predicted multivariate representations as well as \emph{(ii)} their temporal scope 
remain largely unknown.}


Here, we 
\jr{address these issues }by analyzing the brain signals of 304 subjects listening to short stories, while their brain activity was 
recorded with fMRI~\citep{nastase_narratives_2020}. 
\charlotte{First, we confirm that deep language algorithms linearly map onto brain activity~\citep{schrimpf_artificial_2020, caucheteux_disentangling_2021,toneva_interpreting_2019}. Then, we show that adding long-range and hierarchical predictions improves such mapping.}
\alex{After confirming that 
the activations of 
deep language algorithms \jr{linearly} map onto brain activity~\citep{schrimpf_artificial_2020, caucheteux_disentangling_2021,toneva_interpreting_2019}, we show that enhancing these models with long-range and hierarchical predictions improves their \jr{brain mapping.}} 
%
%
%
Critically, \jr{ and in line with predictive coding theory, our} 
results reveal a 
\jr{hierarchical organization of language prediction in the cortex, in which the highest stages forecast \emph{(i)} the most distant and \emph{(ii)} the most abstract representations.}

\begin{figure*}[ht]
\centering
\includegraphics[width=0.95\linewidth]{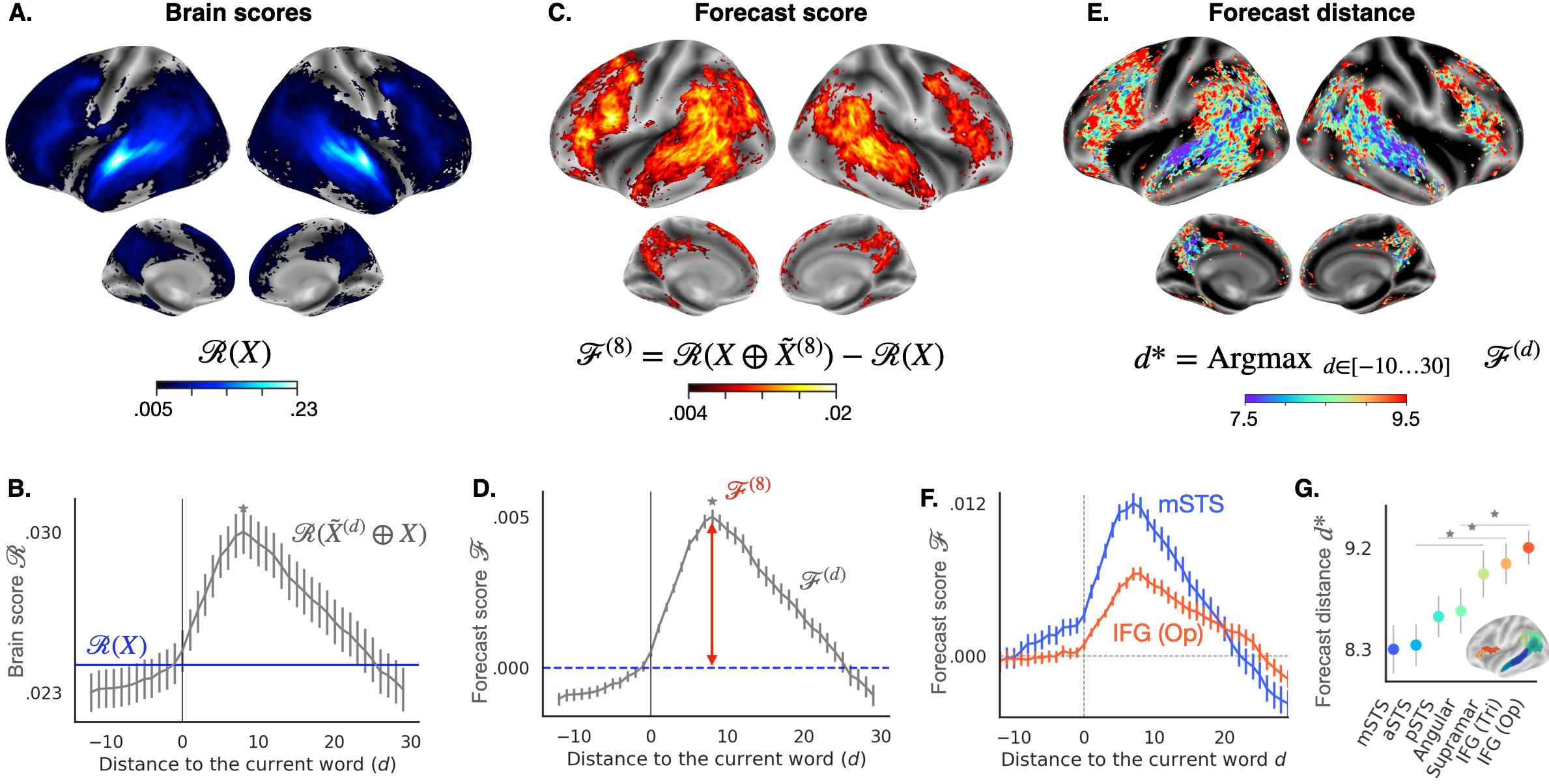}
\caption{
\textbf{Long-range forecasts in the brain.}
\textbf{A.} The ``brain score” ($\mathcal{R}$, Figure \ref{fig:fig1}B, Methods \ref{methods-brainscore}), \jr{obtained with }GPT-2, 
for each subject and each voxel, and \jr{here} 
averaged across subjects (n=304). \charlotte{Only the voxels with significant brain scores are color-coded.}
\textbf{B.} 
\jr{Average (across voxels) brain scores obtained with GPT-2} with (grey) or without (blue) 
forecast representations. \charlotte{The average brain score peaks at $d^*=8$ (grey star).}
\textbf{C.} For each voxel, the \jr{average (across subjects) ``forecast score” $\mathcal{F}^d$, \emph{i.e.} the gain} 
in brain score when concatenating 
\jr{the activations of GPT-2 with a} forecast window $\tilde{X}^{(8)}$.
\charlotte{Only the voxels with significant forecast scores are color-coded.} 
%
\textbf{D.} \jr{Average (across voxels)} forecast scores for different distance $d$. 
\textbf{E.} Distance that maximizes $\mathcal{F}^d$, computed for each subject and each voxel, and denoted $d^*$. This ``forecast distance'' reveals the regions associated with short- and long-range forecasts. 
Regions in red and blue are associated with long-range and short-range forecasts\jr{, respectively}. We only display the voxels with a significant average peak ($\mathcal{F}^{d^*} - \mathcal{F}^{0}, d^*=\,8$, cf. Methods \ref{methods-forecast-distance}).
\textbf{F.} Same as D. but for two selected regions of the brain: the middle superior temporal sulcus (mSTS) and the pars opercularis of the inferior frontal gyrus (IFG-Op). 
\textbf{G.} Forecast distance of seven regions of interest
, as computed for each voxel \jr{of each subject and then} 
averaged 
within the selected brain regions. \jr{For all panels, the error bars are SEM across subjects. All brain maps are thresholded at $p<.01$, as assessed with a FDR-corrected two-sided Wilcoxon test across subjects.}
}
\label{fig:fig2}
\end{figure*}

\begin{figure}[ht]
\centering
\includegraphics[width=\linewidth]{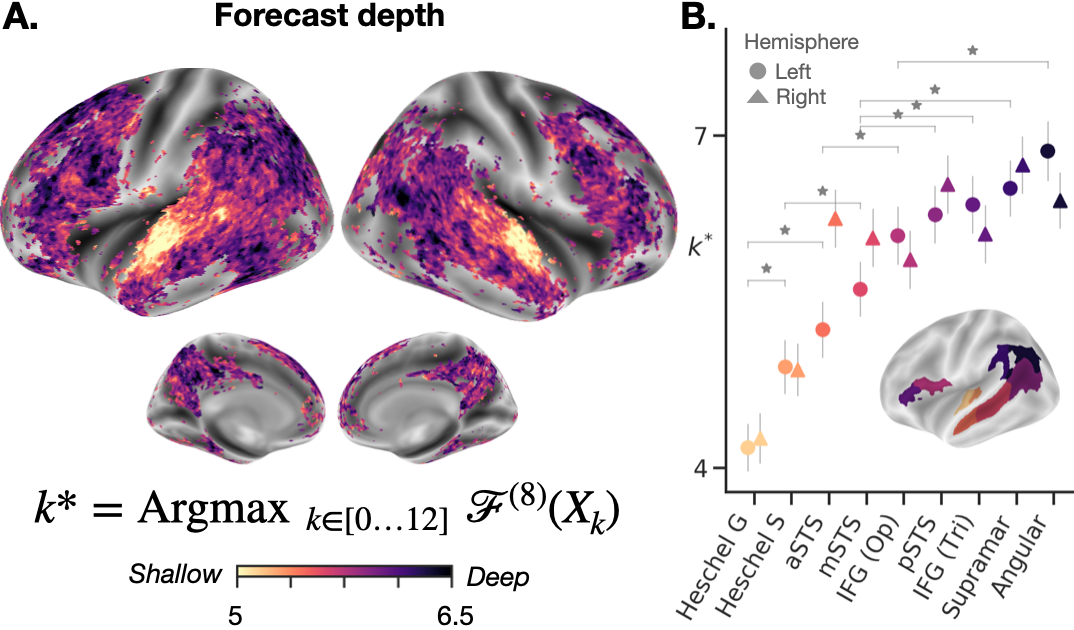}
\caption{\textbf{Hierarchical forecasts in the brain}
\textbf{A.} Depth of the representation that maximizes the forecast score in the brain, denoted $k^*$. Forecast scores are computed for each depth, subject and voxel, at a fix distance $d^*=8$ and averaged across subjects. 
We compute the optimal depth for each subject and voxel \jr{and plot} 
the average forecast depth across subjects. Dark regions are \jr{best} 
accounted for by deep forecasts, while light regions are \jr{best accounted for } 
by shallow forecasts. Only significant voxels are \charlotte{color-coded, following} 
Figure \ref{fig:fig2}C).
\textbf{B.} Same as A, with $k^*$ averaged across the voxels of nine regions of interest, in  
the left (circle) and right (triangle) hemispheres. 
Error bars are SEM across subjects. Pairwise significance between regions is assessed using a two-sided Wilcoxon test on the left hemisphere's scores ($p<.05$). 
}
\label{fig:fig3}
\end{figure}

\begin{figure*}[ht]
\centering
\includegraphics[width=0.7\linewidth]{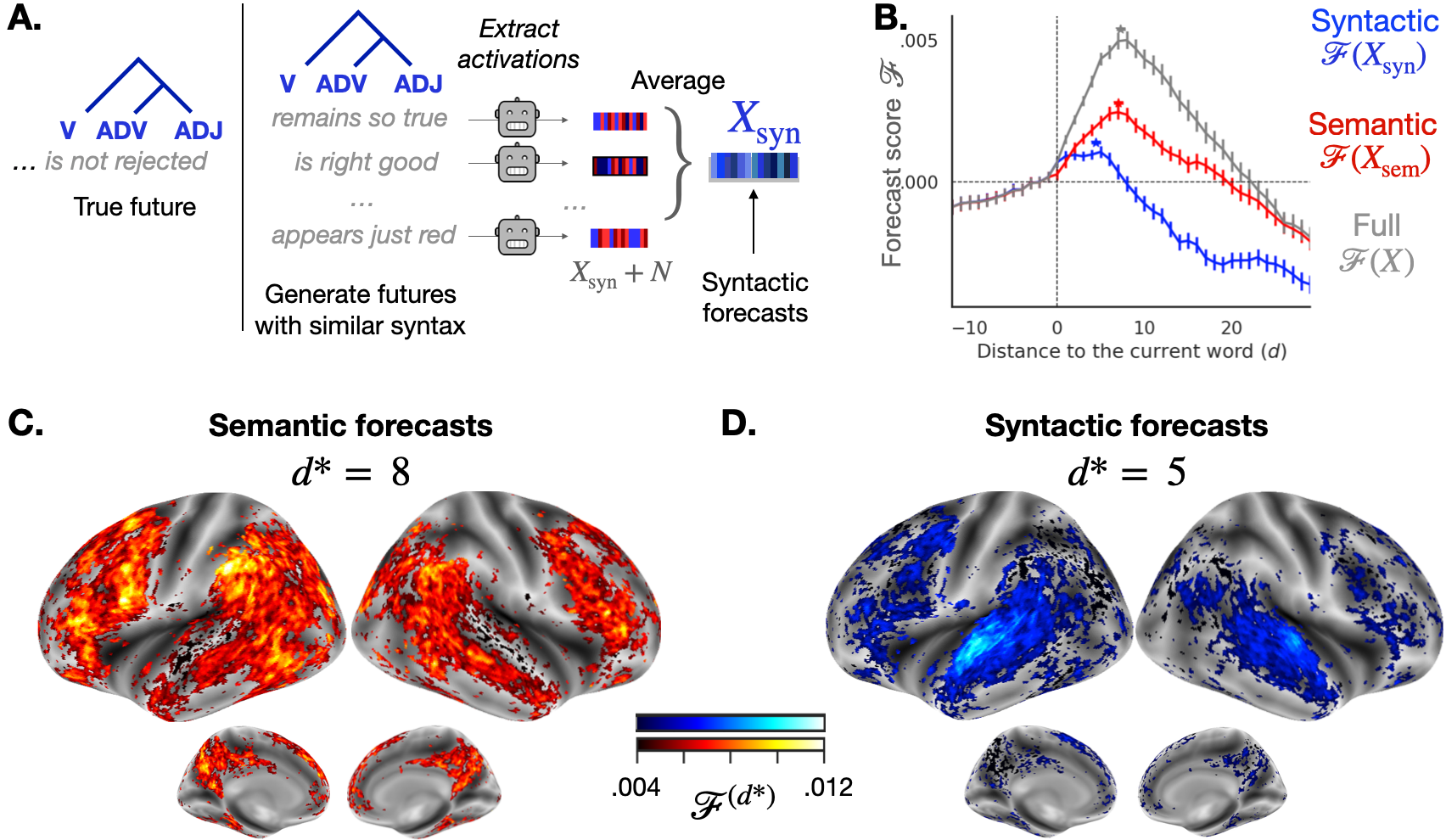}
\caption{\textbf{Syntactic and semantic forecasts in the brain.}
\textbf{A.} Method to extract syntactic and semantic forecast representations,
 adapted from \cite{caucheteux_disentangling_2021}. For each word \jr{and its context} (e.g. `Great, your \textit{paper} ... ', we generate ten possible futures with the same syntax as the original sentence (part-of-speech and dependency tree) but randomly sampled semantics (e.g. `... remains so true'\jr{, `... appears so small'}). Then, we extract the corresponding GPT-2 activations (layer eight). Finally, we average the activations across the ten futures. This method allows to extract the 
syntactic \jr{component common to} 
the ten futures, denoted $X_{\mathrm{syn}}$. 
The semantic \jr{component} 
is \jr{defined as} the residuals of syntax in the full activations; $X_{\mathrm{sem}} = X-X_{\mathrm{syn}}$. We build the syntactic and semantic forecast windows by concatenating the syntactic and semantic components of seven consecutive future words, respectively (Methods \ref{methods-syntactic-forecast-window}). 
\textbf{B.} Syntactic (blue) and semantic (red) forecast scores, \jr{on average across all voxels, following} 
Figure \ref{fig:fig1}C. 
Error bars are SEM across subjects. The average peaks across subjects is indicated with a star.  
\textbf{C.} Semantic forecast scores for each voxel, averaged across subjects and at  $d^*=\,8$, 
\jr{the distance that }maximizes the semantic forecast scores in B. Only significant voxels are displayed \charlotte{similarly to} 
Figure \ref{fig:fig2}C. 
\textbf{D.} Same as C. for syntactic forecast scores and $d^*=\,5$.
}
\label{fig:fig4}
\end{figure*}

\section*{Results}

\paragraph{Deep language models map onto 
brain \jr{activity}.}
First, we 
quantify the 
\jr{similarity} between deep language models and the 
brain\jr{, when these two systems are input with the same stories.} 
For this, we \jr{use the Narratives dataset\footnote{\url{https://openneuro.org/datasets/ds002345/versions/1.1.4.}} \citep{nastase_narratives_2020}, and analyze }the fMRI of 304 subjects listening to \jr{$\approx$70\,min of} short stories. 
\jr{We then fit, for each voxel and each subject independently, a linear ridge regression to predict the fMRI signals }from the activations of 
\jr{a variety of }deep language models. 
%
Finally, we compute the \jr{corresponding} ``brain scores” \jr{using held-out data}, \emph{i.e.} the \jr{voxel-wise} correlation between 
\jr{\emph{(i)}} the fMRI signals and
\jr{\emph{(ii)} the predictions of the ridge regression input with the } activations \jr{of a given language model} 
(Figure \ref{fig:fig1}B).
\jr{For clarity, we first focus on }the activations 
\jr{of the} eighth layer \jr{of} GPT-2 provided by HuggingFace\footnote{ \url{https://huggingface.co/}} \citep{radford_language_nodate}, 
\jr{as} it has been shown to best predict brain activity \citep{schrimpf_artificial_2020,caucheteux_language_2020}. 

In line with previous studies \citep{caucheteux_language_2020,caucheteux_disentangling_2021,wehbe_aligning_2014,jain_incorporating_2018}, the activations of GPT-2 accurately map onto a distributed and bilateral \jr{set of brain areas. }
Brain scores peak in \jr{the auditory cortex, as well as in the anterior temporal and superior temporal areas} (Figure \ref{fig:fig2}A and Figure \ref{fig:rois}). 
%
\jr{The effect sizes of these brain scores are in line with previous work \citep{huth_natural_2016,caucheteux_language_2020,toneva_meaning_2020}: for instance,} the highest brain scores ($R=0.23$, in the superior temporal sulcus (Figure \ref{fig:fig2}A) 
represent $60\,\%$ of the maximum explainable signal, as assessed with a shared-response model across subjects 
(SI.\ref{si-noiseceil} and Figure \ref{fig:noiseceil}). 
Figure \ref{fig:arch} shows that, on average, similar brain scores are achieved with other state-of-the-art language models. 

Overall, these results confirm that deep language models \jr{linearly map onto brain responses to spoken stories.} 

\paragraph{ \jr{Tracking long-range forecast in the brain}.}
We then test whether enhancing \jr{language} 
models with forecast representations 
\jr{leads to higher brain scores (Figure \ref{fig:fig1}D).}
Specifically, for each word, we concatenate \emph{(i)} the model activations of the present word (denoted $X$) and \emph{(ii)} a ``forecast window” \jr{(denoted $\tilde{X}^{(d)}$)}, consisting of the embeddings of future words.
This forecast window is parameterized by a \jr{variable} distance $d$ and a fixed width (seven words, see SI.\ref{si-controls} for the growing window analysis). 
%
%
%
For each distance $d$, we \jr{compute 
the ``forecast score” (denoted $\mathcal{F}^d$) by comparing the 
brain scores obtained with and without the forecast representations 
(Figure \ref{fig:fig2}B).
}

\jr{Our results show that }$\mathcal{F}$ is maximal for a distance of $d=8$ words, and peaks in the areas typically associated with language processing (Figure \ref{fig:fig2}B-D). \jr{These forecast scores are bilaterally distributed in the brain, at the exception of the infero-frontal ($p<\,10^{-5}$ in Pars Opercularis, using a pairwise Wilcoxon test between the left and right hemispheres) and supramarginal gyri ($p<\,10^{-9}$), which exhibit a significant lateralization effect (Figure \ref{fig:rois}B).}
%

Supplementary analyses confirm that \emph{(i)} forecast representations are best \jr{captured} 
with a window size of $\approx$8 words, \emph{(ii)} random forecast representations do not improve the brain scores, and \emph{(iii)} using the words generated by GPT-2 instead of the true future words achieve \jr{lower but} similar 
\jr{results} 
(SI.\ref{si-controls}). 
%

\charlotte{Together, these results reveal long-range 
forecast representations in the brain, which represents a 23\% ($\pm$ 9\% across subjects) improvement in brain \jr{scores} 
(Figure \ref{fig:fig2}AB). }





\paragraph{Forecast distance varies along the cortical hierarchy.}
Do all brain regions predict the same time window?
To \jr{address} 
this issue, we \jr{estimate} 
the peak of the forecast score of each voxel \jr{and} 
denote $d^*$ the corresponding distance. 
The results show that the prefontal areas \jr{forecast}
, on average, \jr{further off} 
in the future than temporal \jr{areas} 
(Figure \ref{fig:fig2}E). 
For instance, 
\jr{$d^*$}
in the inferior temporal gyrus (IFG) 
\jr{is higher than} in the anterior superior temporal sulcus (aSTS) ($\Delta d^* = 0.9 \pm 0.2, p < 10^{-4}$, Figure \ref{fig:fig2}F and G).
\jr{The variation of optimal forecast distance along the temporo-parieto-frontal hierarchy is largely symmetric across the two hemispheres (Figure \ref{fig:rois}).}

\paragraph{\jr{Forecast depth varies along the cortical hierarchy.}}

What is the \textit{nature} 
of these forecast representations? 
To address this issue, we assess whether forecast representations relate to \emph{(i)} 
\jr{shallow or} 
deep 
\jr{as well as} 
\emph{(ii)} syntactic or semantic representations.
%
To this aim, we first compute the forecast scores similarly as in Figure \ref{fig:fig1}C, but \jr{now} 
vary the \jr{layer used} 
from GPT-2. 
Then, we \jr{identify} 
$k^*$ for each voxel \jr{\emph{i.e.} the depth that maximizes} 
the forecast scores (Methods \ref{methods-forecast-depth}).
%

\jr{Our} results show that the optimal forecast depth varies 
\jr{along the expected cortical hierarchy} 
(Figure \ref{fig:fig3}A). Specifically, associative cortices are best modeled with deeper forecasts ($k^*>6$) than low-level language areas (\emph{e.g.} $k^*<6$ in Heschel’s gyri/sulci, anterior STS, Figure \ref{fig:fig3}A-B).
%
%
\charlotte{The difference between regions, while small on average, is highly significant across subjects (e.g.  between the angular and Heschel's giri: $\Delta k^*=2.5 \pm 0.3$, $p<10^{-12}$}), and observed in both the left and right hemispheres (Figure \ref{fig:fig3}B). 

Together, these results suggest that the long-range forecasts of fronto-parietal cortices are more abstract 
than the short-term forecasts of low-level brain regions.

\paragraph{\jr{Short and long range forecasts target syntactic and semantic representations, respectively.}}

\jr{To decompose forecast representations into syntactic and semantic components, }
we apply a method introduced in \citep{caucheteux_disentangling_2021} and proceed as follows: 
for each word \jr{and its preceding context}
, we generate ten possible futures \jr{which matches the syntax of the true future words}
.
\jr{For each of these possible futures, we} extract the corresponding GPT-2 activations, and average 
\jr{them} across the ten possible futures (Figure \ref{fig:fig4}A, Methods \ref{methods-syntactic}). As explained in \citep{caucheteux_disentangling_2021}, this method allows us to \jr{decompose} 
the activations \jr{of a given language model }$X$ into syntactic (the average vector, denoted $X_{\mathrm{syn}}$) and semantic components (the residuals, $X_{\mathrm{sem}} = X-X_{\mathrm{syn}}$). Once the syntactic and semantic forecast windows are built, we compute the corresponding forecast scores (Methods \ref{methods-syntactic-forecast-window}). 

The results show that semantic forecasts are long-range and involve a distributed network peaking in the frontal and parietal lobes. \jr{By contrast, }
syntactic forecasts (Figure \ref{fig:fig4}B) \jr{are relatively short-range and 
localized in the superior temporal 
\jr{and left frontal areas} (Figure \ref{fig:fig4}C and D).}

Overall, these results reveal a hierarchy of predictions in the brain
\jr{in which the superior temporal cortex forecast short-term, shallow and syntactic representations whereas the infero-frontal and parietal areas forecast long-term, abstract and semantic representations.}

\section*{Discussion}





In the present study, we put specific hypotheses of predictive coding theory to the test \citep{rumelhart_interactive_1982,rao_predictive_1999, friston_predictive_2009}:
while deep language algorithms are typically trained to make \emph{(i)} adjacent and \emph{(ii)} word-level  
predictions \citep{vaswani_attention_2017, radford_language_nodate,  devlin_bert_2019, liu_roberta_2019, brown_language_2020, clark_electra_2020}, we \jr{assess whether} 
the human brain predicts \emph{(i)} long-distance and \emph{(ii)} \jr{hierarchical} 
representations. To this aim, we capitalize on the success of a recent methodology \citep{yamins_performance-optimized_2014, khaligh-razavi_deep_2014, guclu_deep_2015, eickenberg_seeing_2017} and compare the activations of
the brain to those of 
\jr{state-of-the-art} deep language 
\jr{models} \citep{huth_natural_2016,jain_incorporating_2018,toneva_interpreting_2019,caucheteux_language_2020,caucheteux_gpt-2s_2021}.
We successfully validate our hypothesis on a large cohort of 304 subjects listening to 70\,min of spoken narratives \citep{nastase_narratives_2020}: brain activity is best explained by deep language algorithms enhanced with long-range and \charlotte{hierarchical forecasts.} 
Our study provides three additional contributions. 

\paragraph{Long-range predictions.}
First, the cortical regions repeatedly linked to high-level semantics, long-term planning, attentional control, abstract thinking and other high-level executive functions \citep{gilbert_executive_2008, shallice_deficits_1991}, namely, the lateral, dorso-lateral and infero-frontal cortices, as well as the supra-marginal gyrus, here exhibit the longest forecast distances.
This result echoes with previous studies showing that the integration constant of the fronto-parietal cortices is larger than those of sensory and temporal areas \citep{wang_dynamic_nodate, lee_anticipation_2021, lerner_topographic_2011, caucheteux_model-based_2021}.
Specifically, our study suggests that these regions, located at the top of the language hierarchy, are not limited to passively integrating past stimuli, but actively anticipate future language representations.

\paragraph{Hierarchical predictions.} 
Second, we show that the depth of forecast representations varies along a similar anatomical organization: the superior temporal sulcus and gyrus are best modeled with low-level forecast representations as compared to the middle temporal, parietal and frontal areas. 
This finding extends previous studies investigating the multiplicity of predictions underlying complex sound or speech processing \citep{vidal_neural_2019, heilbron_tracking_2019, donhauser_two_2020}
\jr{: while previous works focused on correlating brain activity with a subset of hand-crafted and unidimensional prediction \emph{errors} (e.g. word or phoneme surprisal), the present analyses explore, and can thus decompose high-dimensional predictions.}
More generally, our results support \jr{the idea }that, unlike current language algorithms, the brain is not limited to predict word-level representations, \charlotte{but rather makes hierarchical predictions}.

\paragraph{Syntactic and semantic predictions.}
Finally, we use a recent method to decompose \jr{these} neural activations into syntactic and semantic representations \citep{caucheteux_disentangling_2021}, and show that the long-range forecasts 
are predominantly driven by semantic features. 
This finding strengthens the idea that while syntax \jr{may be} 
explicitly represented in neural activity \citep{nelson_neurophysiological_2017,ding2016cortical}, predicting \charlotte{high-level meaning may be at the core of} language processing \citep{jackendoff2002foundations,shain_2021_failure}.

\paragraph{On the potential benefit of a predicting coding architecture.}

Together, these results 
support predictive coding theories, whereby the brain continually predicts sensory inputs, compares these predictions to the truth, and updates its internal model 
accordingly \citep{mcclelland_interactive_1981, rumelhart_interactive_1982, rao_predictive_1999}. 
%
%
Our study further clarifies this general framework:
not only does the brain predict sensory inputs, but each level of the cortical hierarchy appears to be organized to predict different temporal scopes and different levels of abstraction (Figure \ref{fig:fig1}A). 



This computational organization is at odd with current language algorithms which are trained to make adjacent and word-level predictions (Figure \ref{fig:fig1}A). We speculate that the brain architecture \jr{evidenced in this study} presents at least one major benefit over its deep learning counter-parts:
while future observations rapidly become indeterminate in their original format, their latent \alex{and abstract} representations 
\alex{may remain predictable}
over long time periods. This issue is already pervasive in speech- and image-based algorithms and has been partially bypassed
with losses based on pretrained embedding \citep{szegedy_going_2015}, contrastive learning and, more generally,
joint embedding architectures \citep{chen_simple_2020, he_momentum_2020, el-nouby_xcit_2021, bardes2021vicreg}. Here, we highlight that this issue also prevails in \jr{language models}, 
where word sequences -- but arguably not their meaning -- rapidly become \alex{unpredictable.}
Our results suggests that predicting hierarchical representations over multiple temporal scopes may be critical to address 
the indeterminate nature of \jr{such} distant observations.


Beyond \jr{clarifying the  }
the brain \jr{and computational }bases of language
, our study thus calls for training 
algorithms 
\jr{to predict a hierarchical representation of future inputs.}



\matmethods{


\subsection{Notations}
We denote:

\begin{itemize}
\item $w$ a sequence of $M$ words (here, several short stories).
\item $X$ the 
activations \jr{of a deep language model input with} $w$, of size $M \times U$, with $U$ the dimensionality of the embeddings (
\jr{for a layer of GPT-2}, $U=\,768$). 
\jr{Except if stated otherwise}, we use the activations extracted from the eighth layer of \jr{a 12-layer} 
GPT-2 \jr{model} (Methods \ref{methods-dlm}). We will explicitly denote $X_k$ the activations extracted from layer $k$ when using another layer. 
\item $Y$ the fMRI recordings elicited by $w$, of size $T \times V$, with $T$ the number of fMRI time samples, and $V$ the number of voxels (Methods \ref{methods-fmri}).
\item $\mathcal{R}(X)$ the brain score of $X$ 
(Methods \ref{methods-brainscore}).
\item $\widetilde{X}^{(d)}$ the forecast window containing information up to $d$ words in the future. In short, the forecast window is the concatenation of 
\charlotte{the deep net activations of seven successive words}, the last word being at a distance $d$ from the current word (Methods \ref{methods-forecast-window}). 
\item $\mathcal{F}^{(d)}(X)$, the forecast score at distance $d$, i.e. the \jr{gain} 
in brain score when concatenating the forecast window $\tilde{X}^{(d)}$ \jr{to the network's activations}; $\mathcal{F}^{(d)}(X) = \mathcal{R}(X \oplus \tilde{X}^{(d)}) - \mathcal{R}(X)$ (Methods \ref{methods-forecast-distance}).
\item $d^*$, the distance maximizing the forecast score; $d^* = \mathrm{argmax}_{d \in [-10, \dots, 30]} \ \mathcal{F}^{(d)}(X)$ (Methods \ref{methods-forecast-distance}).
\item $k^*$, the network's depth maximizing the forecast score at a fixed distance $d=8$; $k^* = \mathrm{argmax}_{k \in [0, \dots, 12]} \ \mathcal{F}^{(8)}(X_k)$, with $X_k$ the activations extracted from the $k^{\mathrm{th}}$ layer of GPT-2. We use $d=\,8$ because it is the distance with the best forecast score on average across subjects and voxels (Methods \ref{methods-forecast-depth}). 
\end{itemize}

\subsection{fMRI dataset}\label{methods-fmri} 
We use the brain recordings (denoted $Y$) of the ``Narratives'' dataset \citep{nastase_narratives_2020}, a publicly available dataset containing the fMRI recordings of 345 subjects listening to 27 spoken stories in English, from $\approx$3\,min to $\approx$56\,min ($\approx$4.6\,h of unique stimulus in total). We use the 
pre-processed fMRI \jr{signals} from the original dataset, without spatial smoothing (referred to as ``afni-nosmooth'' in the repository) \jr{and sampled with TR=1.5\,s}: the preprocessing steps were performed using fMRIPrep \citep{esteban_fmriprep_2019}, no temporal filtering was applied.\jr{The resulting preprocessing leads to the analysis of cortical voxels projected onto the surface and morphed onto a "fsaverage" template brain, and hereafter referred to as voxels simplicity}. As suggested in the original paper, 
\jr{some} subject-story pairs were excluded because of \jr{noise}
, resulting in 622 subject-story pairs and 4\,h of \jr{unique} audio \jr{material} in total. 

\subsection{Deep language models' activations}\label{methods-dlm} 
We compare the fMRI recordings with the activations of \jr{a variety of pretrained }deep language model 
input with the same sentences presented to the subjects.  
\jr{For clarity, we primarily focus on GPT-2,} a high-performing language model trained 
to predict 
words given 
\jr{their} previous context. 
\charlotte{GPT-2 consists} of twelve Transformer modules \citep{vaswani_attention_2017, radford_language_nodate}, each of them 
\jr{referred to as} ``layer'', stacked onto one non-contextual word embedding layer. Here, we use the pre-trained models from Huggingface \cite{wolf-etal-2020-transformers} (1.5 billion parameters, trained on 8 million web pages) \jr{(\emph{c.f.} SI \ref{si-controls} for the other deep language models).}

In practice, to extract the activations $X$ elicited by a sequence of $M$ words $w$, from the $k^{th}$ layer of the network, we 1) format the textual transcript of the sequence $w$ (replacing special punctuation marks such as "–" and duplicated marks "?." by dots) 2) tokenize the text using Huggingface tokenizer, 3) input the network with the tokens and 4) extract the corresponding activations from layer $k$. This results in a vector of size $M \times U$, with $M$ the number of words and $U$ the number of units per layer (here $U=\,768$). Given the constrained context size of the network, each word is successively input to the network with at most 1024 previous tokens. For instance, while the third word's vector is computed by inputting the network with $(w_1, w_2, w_3)$, the last word's vectors $w_M$ is computed by inputting the network with $(w_{M-1024}, \dots, w_M)$. The alignment between the stories' audio recordings and their textual transcripts was provided in the original Narratives database \citep{nastase_narratives_2020}. 


\subsection{Brain scores
}\label{methods-brainscore} 

Following previous works \cite{huth_natural_2016,caucheteux_language_2020,caucheteux_gpt-2s_2021}, we evaluate, for each subject $s$ and voxel $v$, the mapping between 1) the fMRI activations $Y^{(s, v)}$ in response to the \jr{audio-}stories 
and 2) the activations $X$ of \jr{the deep network  input with }
the \charlotte{textual transcripts of the} same stories. To this end, we fit a linear \jr{ridge regression} 
$W$ on a train set to predict the fMRI scans given the \jr{network's activations.} 
Then, \jr{we evaluate this }
mapping 
by computing the Pearson correlation between predicted and actual fMRI scans on a held out set:
\begin{equation}\label{equ_mapping}
\mathcal{R}^{(s,v)} : X \mapsto  \mathrm{Corr} 
\big( 
 W \cdot X, Y^{(s,v)}
  \big) \quad , 
\end{equation}
with $W$ the fitted linear projection, $\mathrm{Corr}$ Pearson's correlation, $X$ the activations of GPT\nobreakdash-2 and $Y^{(s, v)}$ the fMRI scans of one subject $s$ at one voxel $v$, both elicited by the same held out stories. 


In practice and following \cite{huth_natural_2016}, we \jr{model} 
the slow bold response \jr{thanks to }
a finite impulse response (FIR) 
model with 6 delays. 
\jr{Still following \cite{huth_natural_2016}, we sum the model activations of the words presented within the same TR, in order to match the sampling frequency of the fMRI and the language models.}
Then, we estimate the linear mapping $W$ with a $\ell_2$-penalized linear regression after standardizing the data, and reducing their dimensionality (computational reasons). 
%
We follow scikit-learn implementation \cite{scikit-learn} and use a pipeline with the following steps: standardization of the features (set to 0 mean with a standard deviation of 1 using a `StandardScaler'), principal component analysis (PCA) with twenty components\footnote{Twenty is the number of components corresponding to the ``elbow'' of the explained variance ratio when applying principal component analysis on the activations of the eighth layer of GPT-2, `RidgeCV' model cross the words of the stories.},  $\ell_2$-penalized linear regression (`RidgeCV' in scikit-learn). The regularization hyperparameter of the `RidgeCV' is selected with \jr{a nested} leave-one-out cross-validation among ten possible values log-spaced between $10^{-1}$ and $10^{8}$ for each voxel and each training fold. 

The outer cross-validation scheme allowing for an independent performance evaluation, uses five folds obtained by splitting the fMRI time series into five contiguous chunks.
%
The Pearson correlations averaged across the five test folds is called ``brain score'', denoted $\mathcal{R}^{(s,v)}(X)$. It measures the mapping between the activation space $X$ and the brain of one subject $s$ at one voxel $v$, in response to the same language stimulus. 

In Figure \ref{fig:fig2}A and B, brain scores are computed for each (subject, voxel) pair. We then average the brain scores across subjects (Figure \ref{fig:fig2}A) and/or voxels (Figure \ref{fig:fig2}B) depending on the analysis. For simplicity, we denote $\mathcal{R}(X)$ the brain scores averaged across subjects and/or voxels.

\subsection{Forecast windows}\label{methods-forecast-window}
The forecast window at distance $d$, denoted $\widetilde{X}^{(d)}$, is the concatenation of the network's activations of seven successive words, the last one being at a distance $d$ from the current word. Precisely, the forecast window of a word $w_n$, at a distance $d$ is the concatenation of the network's activations elicited by words $w_{n+d-7}, \dots, w_{n+d}$. 
Thus, 
\begin{equation}\label{equ_window}
\widetilde{X}^{(d)}  = ( X_{w_{n+d-7}} \oplus  \dots \oplus X_{w_{n+d}} )_{n\in [1, \dots, M]} \quad , 
\end{equation}
with $\oplus$ the concatenation operator, and $M$ the number of words in the transcript $w$. Note that $d$ can be negative: in that case, the forecast window only contains past information. 
\jr{Except if stated otherwise, }the forecast window is built out of the activations $X$ extracted from the eighth layer of GPT-2. In Figure \ref{fig:fig3}, the forecast window is built out of the activations $X_k$  extracted from different layers $k$ of GPT-2. We denote $\widetilde{X}^{(d)}_k$ the corresponding forecast windows. In Figure \ref{fig:fig4}, the forecast windows are built out of the syntactic ($X_{\mathrm{syn}}$) and semantic  ($X_{\mathrm{sem}}$) activations of GPT-2 (cf. Methods \ref{methods-syntactic} and \ref{methods-syntactic-forecast-window}). 

\subsection{Forecast scores}\label{methods-forecast-score}
For each distance $d$, subject $s$ and voxel $v$, we compute the ``forecast score'' $\mathcal{F}^{(d, s, v)}$, which is the \jr{gain} 
in brain score when concatenating the forecast windows to the present GPT-2 activations. Thus, 
\begin{equation}\label{equ_forecast}
\mathcal{F}^{(d, s, v)}: X \mapsto \mathcal{R}^{(s, v)}(X \oplus \widetilde{X}^{(d)}) - \mathcal{R}(X) \quad ,
\end{equation}

\charlotte{To match the dimensionality of $X$ and $\tilde{X}$, the principal component analysis used to compute the mapping (Methods \ref{methods-brainscore}) was trained on $X$ and $\tilde{X}$ separately, before concatenating the two features: i.e. $\mathcal{F}(X) = \mathcal{R}(\mathrm{pca}(X)+\mathrm{pca}(\tilde{X})) - \mathcal{R}(\mathrm{pca}(X))$.}
%


\subsection{Forecast distance}\label{methods-forecast-distance}
To test whether the forecast scope varies along the cortical hierarchy, 
we 
\jr{estimate} the distance that maximizes the forecast score. Precisely, the optimal ``forecast distance'' $d^*$ for each subject $s$ and voxel $v$ is 
\charlotte{defined as:}
%
\begin{equation}\label{equ_distance}
d^{*}_{(s, v)} = \mathrm{argmax}_{d \in [-10, \dots, 30]} \ \mathcal{F}^{(d, s, v)}(X) \enspace ,
\end{equation}
with $X$ the 
activations \jr{of the language model}, $\mathcal{F}^{(d, s, v)}$ the forecast score at distance $d$ for subject $s$ and voxel $v$ (\eqref{equ_forecast}). The forecast distances $d^*$ are then averaged across subjects and/or voxels depending on the analyses. 

The present analysis is only relevant for the brain regions for which forecast scores are not flat. 
Indeed, computing the distance maximizing a flat curve would be misleading. Thus, in Figure \ref{fig:fig2}E, we 
compute the difference $\mathcal{F}^{8} - \mathcal{F}^{0}$ for each subject and voxel, assess the significance with Wilcoxon test across subjects, and ignore the voxels with a non-significant difference ($p>\,.01$).  

\subsection{Forecast’s depth}\label{methods-forecast-depth}

To test whether the forecast depth varies along the cortical hierarchy, 
we compute the forecast score for different depth of representation. Precisely, we proceed similarly as in \ref{methods-forecast-score}, but replacing $X$ 
\charlotte{by }the activations $X_k$ extracted from layer $k$ of GPT-2 ($k \in [0, \dots, 12]$) in \eqref{equ_forecast} and \eqref{equ_window}. Then, we compute
%
%
%
the depth maximizing the forecast score, called ``forecast depth'', and given by:
\begin{equation}
k^*_{(d, s, v)} = \mathrm{argmax}_{k \in [0, \dots, 12]}  \mathcal{F}^{(d, s, v)}(X_k) \quad , 
\end{equation}
\charlotte{with $\mathcal{F}^{(d, s, v)}(X_k)  = \mathcal{R}^{(s, v)}(X_k \oplus \widetilde{X_k}^{(d)}) - \mathcal{R}(X_k)$ (\eqref{equ_forecast}).}
For simplicity, 
we focus on the fixed distance $d=\,8$ (Figure \ref{fig:fig3}C and D), 
\jr{which} maximizes the forecast score in Figure \ref{fig:fig2}. 


\subsection{\jr{Decomposing the activations of language models }
into syntactic and semantic components}\label{methods-syntactic}
To extract the syntactic and semantic components of $X$, \jr{a vector of activations in response to a story $w$}, we apply a method introduced in \citep{caucheteux_disentangling_2021} (Figure \ref{fig:fig4}A). For each word, 1) we generate $k=10$ futures of the same syntax as the true future (\emph{i.e.} same part-of-speech and dependency tags as the true future), but randomly sampled semantics, 2) we compute the activations for each of the ten possible futures,
and 3) we average the activations across the ten futures. 
This method allows to extract the average vector $X_{\mathrm{syn}}$, that contains syntactic information but is deprived from semantic information. The semantic activations $X_\mathrm{sem} = X- X_\mathrm{syn}$ are the residuals of syntax in the full activations $X$. 

\subsection{Syntactic and semantic forecast windows}\label{methods-syntactic-forecast-window}
To investigate syntactic and semantic forecasts in the brain, we build forecast windows out of the syntactic and semantic activations of GPT-2, respectively. To this aim, we first build the forecast windows out of GPT-2 activations $\widetilde{X}^{(d)}$, similarly as \ref{methods-forecast-window}. Then, we extract the syntactic $\widetilde{X}^{(d)}_\mathrm{syn}$ and semantic $\widetilde{X}^{(d)}_\mathrm{sem}$ components of the concatenated activations, as introduced in \cite{caucheteux_disentangling_2021} and described in \ref{methods-syntactic}. Finally, the syntactic forecast score is the increase in brain score when concatenating the syntactic window:  

\begin{equation}
\mathcal{F}^{(d)}_\mathrm{syn} = \mathcal{R} (X \oplus \widetilde{X}^{(d)}_\mathrm{syn} ) - \mathcal{R}(X)
\end{equation}
Similarly, the semantic forecast score is given by: 

\begin{equation}
    \mathcal{F}^{(d)}_\mathrm{sem} = \mathcal{R}(X \oplus \widetilde{X}^{(d)}_\mathrm{sem}) - \mathcal{R}(X)
\end{equation}

\subsection{Brain parcellation}\label{methods-parcellation}
We systematically \jr{implement} 
whole brain analyses and compute scores for each voxel in the brain. Yet, for simplicity, we report the scores averaged across selected regions of interest in Figure \ref{fig:fig2}F,G  and \ref{fig:fig3}C. To this aim, we use a subdivision of the Destrieux Atlas \cite{destrieux_automatic_2010}. Regions with more than 500 vertices are split into smaller parts. This results in 142 regions per hemisphere, each containing less than 500 vertices. In Figure \ref{fig:fig2}G and \ref{fig:fig3}C, we use the following acronyms:

\begin{table}[h]
\centering
\begin{tabular}{lr}
Acronym & Definition \\
\midrule
STG / STS & Superior temporal gyrus / sulcus \\
aSTS  & Anterior STS \\
mSTS  & Mid STS \\
pSTS  & Posterior STS \\
Angular / Supramar & Angular / Supramarginal inferior parietal gyrus \\
IFG / IFS & Inferior frontal gyrus / sulcus\\
Tri / Op & Pars triangularis / opercularis (IFG)\\
Heschel G / Heschel S & Heschel gyrus / sulcus \\
\bottomrule
\end{tabular}
\end{table}

\subsection{Statistical significance}\label{methods-significance}
We systematically \jr{implement} 
single-subject and whole brain analyses: all metrics (brain score, forecast score, forecast distance and depth) are computed for each subject, voxel pair. We report the metrics averaged across subjects and/or voxels depending on the analysis. Statistics are computed across subjects, using the two-sided Wilcoxon test from Scipy  \cite{2020SciPy-NMeth} assessing whether the metric (or the difference between two metrics) is significantly different from zero. We \jr{report} 
an effect \jr{as} 
significant \jr{if its }
p-value is lower than 0.01. Error bars systematically refer to the Standard Errors of the Means (SEM) across subjects, following Scipy implementation.
}

\showmatmethods{}

\section*{References}
\bibliography{references}

\newpage
\quad
\newpage
\section*{Supporting Information (SI)}
\renewcommand{\thefigure}{S\arabic{figure}}

\subsection{Generalisation to other architectures}

\begin{figure}[ht]
\centering
\includegraphics[width=\linewidth]{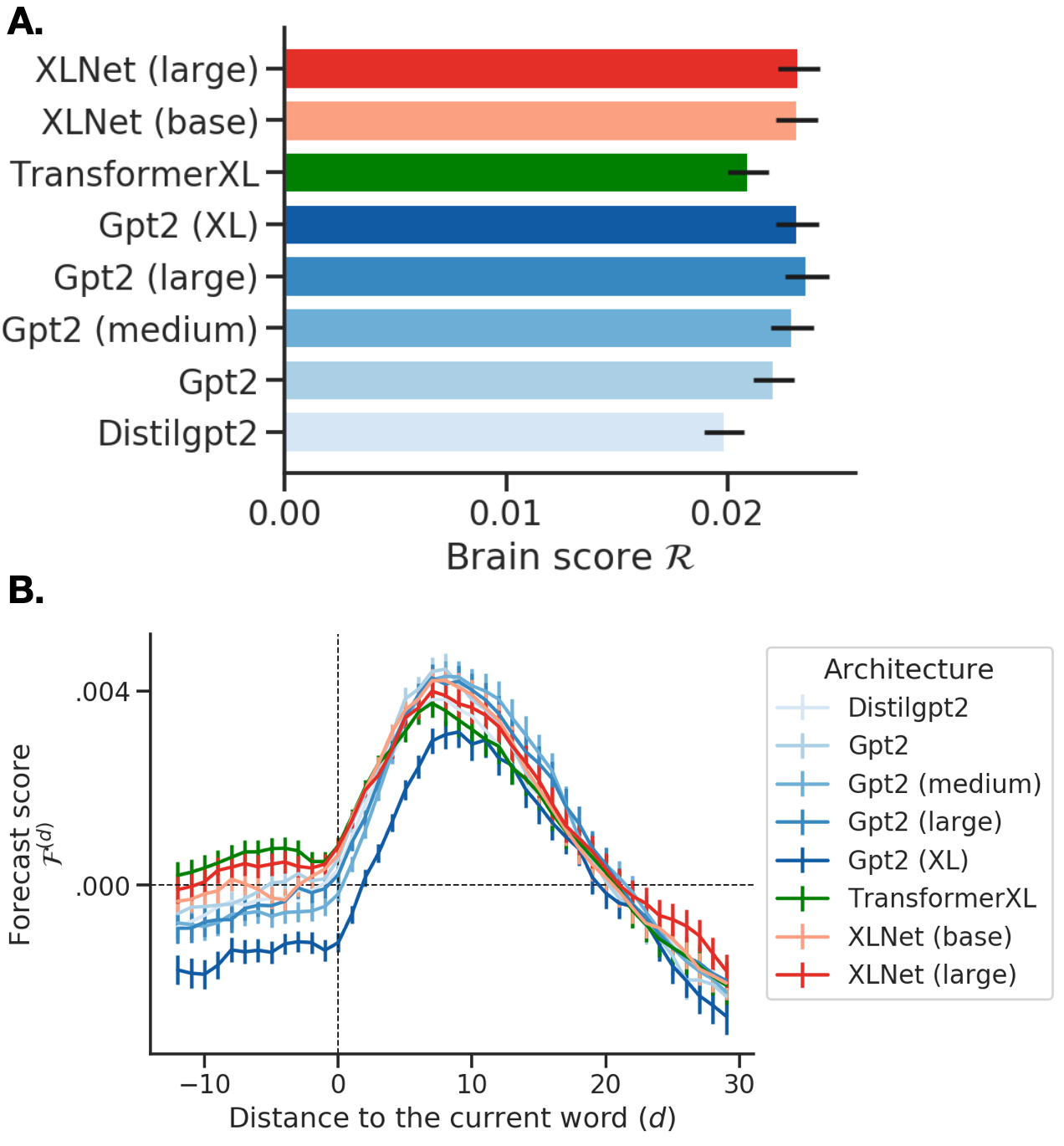}
\caption{\textbf{Generalisation to other architectures.} \textbf{A.} Brain scores (cf. Figure \ref{fig:fig1}B, Methods \ref{methods-brainscore}) of eight transformer models, based on XLNet \cite{yang_xlnet_2020}, TransformerXL \cite{dai_transformer-xl_2019} and GPT-2 \cite{radford_language_nodate} architectures. We use the pre-trained models from Huggingface and proceed similarly as with GPT-2 (Methods \ref{methods-dlm}). Brain scores are averaged across voxels and subjects, error bars are the standard errors of the mean across subjects. \textbf{B.} Same as Figure \ref{fig:fig2}D for the eight transformer architectures. 
}
\label{fig:arch}
\end{figure}

The analyses in the main manuscript focus on one representative deep neural network: GPT-2 \cite{radford_language_nodate}. Here, we replicate our results with the activations extracted from seven other transformer architectures. We only analyse \textit{causal} models, trained to predict a word from their \textit{previous} context\footnote{Note that XLNet is trained to predict both left and right context \citep{yang_xlnet_2020}, but, here, we only input the model with left context when extracting the activations.}. Similarly as with GPT-2, we use the pretrained models from Huggingface (labeled `distilgpt2', `gpt2', `gpt2-medium', `gpt2-large', `gpt2-large', `gpt2-xl', `transfo-xl-wt103', `xlnet-base-cased', `xlnet-large-cased'), based on GPT-2 \citep{radford_language_nodate}, XLNet \citep{yang_xlnet_2020} and Transformer-XL \citep{dai_transformer-xl_2019} architectures, and focus on one intermediate-to-deep layer of the model ($l = \frac{2}{3} \times n_{\mathrm{layers}}$). For each architecture, we 1) extract the activations corresponding to the subjects' stories (Methods \ref{methods-dlm}) 2) compute the corresponding brain scores (Methods \ref{methods-brainscore}) and forecast scores (Methods \ref{methods-forecast-score}) for each voxel, subject, and forecast distance. As displayed in Figure \ref{fig:arch}, the seven architectures accurately map onto brain activity (Figure \ref{fig:arch}A), and the mapping is improved when adding information about around eight words in the future (Figure \ref{fig:arch}B). \charlotte{The mapping is also improved when adding representations of words automatically generated by GPT-2 instead of the true future words (we use sampling methods to generate words, similarly as in SI.\ref{si-controls}).} 

\subsection{Controls} \label{si-controls}

\begin{figure}[ht]
\centering
\includegraphics[width=0.8\linewidth]{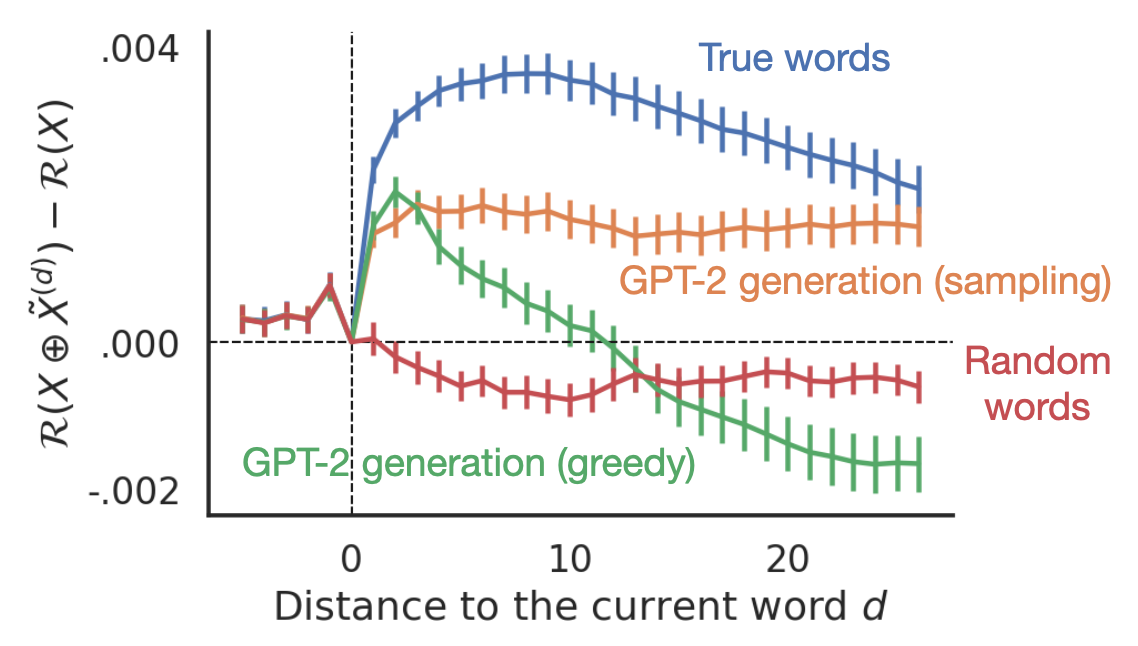}
\caption{\textbf{Controls.} Forecast scores for different types of forecast representations $\widetilde{X}$. Here, we use a growing window analysis: $\widetilde{X}^{(d)}$ is the concatenation of the activations of $|d|$ future ($d>0$) or past ($d<0$) words; the size of the window thus varies with the distance. The forecast score is the gain in brain score when concatenating the forecast window (cf. \eqref{equ_forecast}). In blue, $\widetilde{X}$ is built out of the true words of the story. In red, $\widetilde{X}$ is built out of randomly picked words from all stories. In green and orange, $\widetilde{X}$ is built out of words generated by GPT-2. Precisely, GPT-2 is input with the current word and its previous context, and we use greedy (green) and sampling (orange) decoding schemes to generate a sequence of expected words. 
For simplicity, when $d<0$, $\widetilde{X}$ is the concatenation of $d$ the \textit{true} past words. When $d>0$, $\widetilde{X}$ is the concatenation of $d$ future words (either true, generated or random words). }
\label{fig:controls}
\end{figure}

\paragraph{Testing different window sizes}

In the previous paragraphs, 
we use a sliding forecast window with a \textit{fixed} number of words in order to compare the brain scores of representations with the same dimensionality. Here, we test different window sizes by 
\jr{implementing} a growing window analysis. Precisely, we build the forecast window $\tilde{X}^{(d)}$ by concatenating the $d$ words succeeding the current word. The size of the window thus varies and $d$ corresponds to both the number of words in the window, and the distance between the last word and the current word. We proceed similarly as in the main manuscript, build forecast window for different distances $d$ and the corresponding forecast scores. As displayed in Figure \ref{fig:controls}, the forecast score is maximal for a window of 8 future words ($d^*= 7.9 \pm 0.5$ on average across subjects), which is consistent with the previous results (Figure \ref{fig:fig2}C, where $d^*=8$).


\paragraph{Using random forecast representations}

We use the same growing window framework and check that adding a forecast window composed of random words does not improve the brain score (Figure \ref{fig:controls}). Precisely, we randomly pick words out of all stories, concatenate the GPT-2 activations of random words to build the forecast windows $\tilde{X}^{(d)}$, and compute the corresponding forecast scores for different distances $d$. Figure \ref{fig:controls} shows that random forecast windows do \textit{not} improve our ability to predict brain activity. 

\paragraph{Using GPT-2 generations as forecast representations}
To what extent are the improvements in brain score due to (1) additional information about future words and/or (2) a different way to represent past words?
To address this question, 
we repeat the same analysis with a forecast window input, not of the \textit{true} future words, but with the words \textit{generated} by GPT-2.
Specifically, for each word $w_k$, we 1) \jr{input} 
GPT-2 with its past context $w_0, \dots w_k$, 2) generate future words $w'_{k+1}, \dots w'_{k+n}$ using different decoding methods (greedy and sampling schemes\footnote{Using Huggingface's sampling scheme with topk$=$50 and topp$=$0.95, do\_sample$=$True, max\_length$=$100. For the greedy scheme, we simply set do\_sample to False, topp and topk to 1.} \citep{holtzman_curious_2020}), 3) extract the corresponding activations $X'_{k+1}, \dots X'_{k+n}$, 4) build the growing windows from these activations and 5) compute their forecast scores. 
Thus, the brain signals, the current activations $X_k$ and the activations of generated words $X'_{k+n} \dots X'_{k+n}$ are all distinct transformations of the same past words $w_0, \dots w_k$. The results show that a window made of \textit{generated} words improves the brain score, although less \jr{so} than a window made of the \textit{true} words of the stories (Figure \ref{fig:controls}), confirming that GPT-2 is an imperfect forecaster.

\subsection{Shared-Response-Model 
\jr{\emph{a.k.a}} ``Noise ceiling''}\label{si-noiseceil}
\begin{figure}[ht]
\centering
\includegraphics[width=0.6\linewidth]{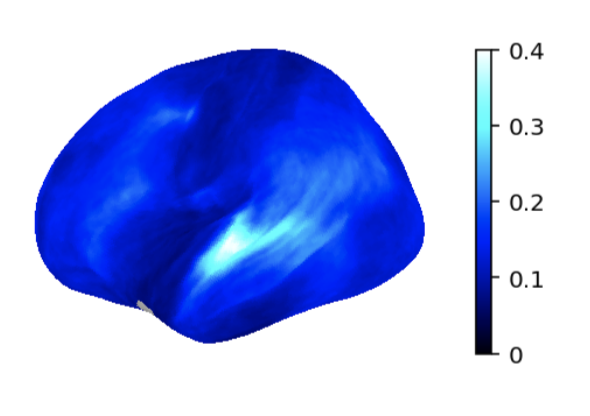}
\caption{\textbf{
\jr{Shared response model}
} Noise ceiling estimates averaged across subjects, for each voxels of the left hemisphere (SI. \ref{si-noiseceil}).
}
\label{fig:noiseceil}
\end{figure}

FMRI recordings are inherently noisy. To assess the amount of explainable signal, we use a shared-response-model, \textit{i.e.} we predict the brain responses of one subject given the other subjects' responses to the same story. Precisely, for one subject $s$ and voxel $v$, we apply the exact same setting as \eqref{equ_mapping}, but use the average brain signals of other subjects' brain $\overline{Y}^{(s)} = \frac{1}{|\mathcal{S}|} \sum_{s' \neq s} Y^{(s')}$ (of size $T \times V$) instead of the network's activations $X$.
Thus, the `noise ceiling' of one subject $s$ and voxel $v$ evaluated on one test set $I$, is given by:
\begin{equation}\label{equ_noiseceil}
 \mathrm{Corr} 
\big( 
 W \cdot \overline{Y}^{(s)}, 
  Y^{(s,v)}
  \big) \quad , 
\end{equation}
with $\mathrm{Corr}$ Pearson's correlation and $W$ a $\ell_2$-penalized linear regression fitted on separate train data, following the notation of \eqref{equ_mapping}. The noise ceilings are computed on five test folds using a cross-validation scheme across time samples and averaged across the five test folds, following the exact same framework as in Methods \ref{methods-brainscore}. Results are displayed in Figure \ref{fig:noiseceil}. This score is one possible upper bound for the best brain score that can be obtained given the level of noise in the dataset.

\subsection{Scores per region of interest}
\begin{figure*}[ht]
\centering
\includegraphics[width=0.7\linewidth]{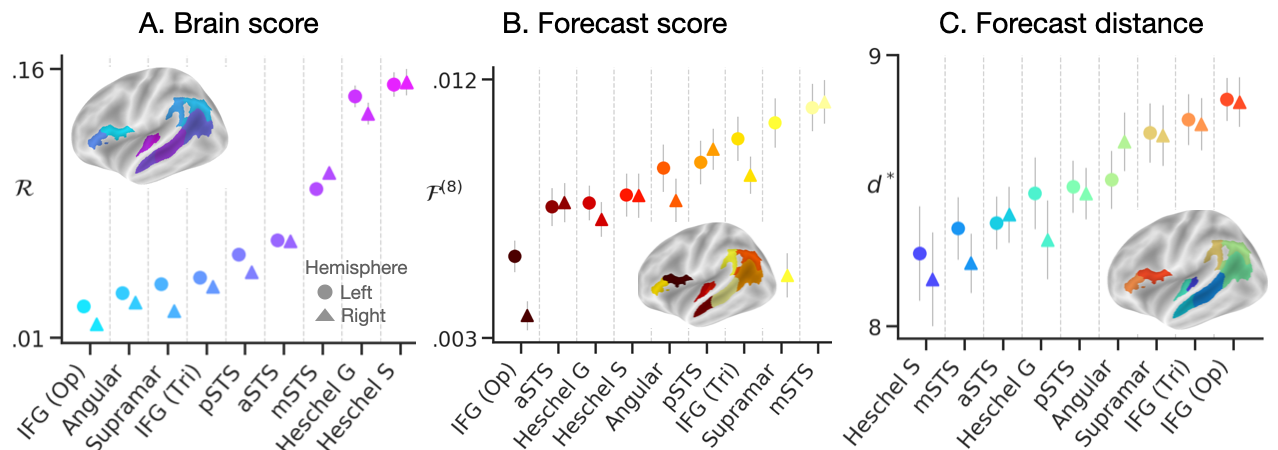}
\caption{\textbf{Scores per region of interest.} \textbf{A-E.} Brain scores (Figure \ref{fig:fig2}A, Methods \ref{methods-brainscore}), forecast scores (Figure \ref{fig:fig2}C, Methods \ref{methods-forecast-score}),  forecast distance (Figure \ref{fig:fig2}E Methods \ref{methods-forecast-distance}) and forecast depth (Figure \ref{fig:fig3}A, Methods \ref{methods-forecast-depth}) for nine regions of interests in both the left (circle) and right (triangle) hemispheres. Scores are averaged across voxels within each region of interest and across subjects. Error bars are the standard errors of the mean across subjects. Regions are ordered with respect to their average score in the left hemisphere.}
\label{fig:rois}
\end{figure*}

For clarity, we report below (Figure \ref{fig:rois}) the average brain scores, forecast scores, forecast distances and depths for each region of interest in both the left and right hemispheres.

\end{document}